\documentclass[sigchi]{acmart}

\renewcommand\footnotetextcopyrightpermission[1]{} %
\usepackage{xurl}
\usepackage{makecell}
\usepackage{enumitem}
\usepackage{subfigure}
\AtBeginDocument{%
  \providecommand\BibTeX{{%
    \normalfont B\kern-0.5em{\scshape i\kern-0.25em b}\kern-0.8em\TeX}}}

\begin{document}
\settopmatter{printacmref=false} %

\title{365 Dots in 2019: \\Quantifying Attention of News Sources}
\author{Alexander C. Nwala}
\affiliation{%
  \institution{Old Dominion University}
  \city{Norfolk} 
  \state{Virginia} 
  \postcode{23529}
  \country{USA}
}
\email{anwala@cs.odu.edu}

\author{Michele C. Weigle}
\affiliation{%
  \institution{Old Dominion University}
  \city{Norfolk} 
  \state{Virginia} 
  \postcode{23529}
  \country{USA}
}
\email{mweigle@cs.odu.edu}

\author{Michael L. Nelson}
\affiliation{%
  \institution{Old Dominion University}
  \city{Norfolk} 
  \state{Virginia} 
  \postcode{23529}
  \country{USA}
}
\email{mln@cs.odu.edu}

\renewcommand{\shortauthors}{Nwala, Weigle, and Nelson}

\pagestyle{empty}
\begin{abstract}
We investigate the overlap of topics of online news articles from a variety of sources. To do this, we provide a platform for studying the news by measuring this overlap and scoring news stories according to the degree of attention in near-real time. This can enable multiple studies, including identifying topics that receive the most attention from news organizations and identifying slow news days versus major news days.  Our application, StoryGraph, periodically (10-minute intervals) extracts the first five news articles  from the RSS feeds of 17 US news media organizations across the partisanship spectrum (left, center, and right). From these articles, StoryGraph extracts named entities (\texttt{PEOPLE}, \texttt{LOCATIONS}, \texttt{ORGANIZATIONS}, etc.) and then represents each news article with its set of extracted named entities. Finally, StoryGraph generates a news similarity graph where the nodes represent news articles, and an edge between a pair of nodes represents a high degree of similarity between the nodes (similar news stories). Each news story within the news similarity graph is assigned an attention score which quantifies the amount of attention the topics in the news story receive collectively from the news media organizations. The StoryGraph service has been running since August 2017, and using this method, we determined that the top news story of 2018 was the \textit{Kavanaugh hearings} with attention score of 25.85 on September 27, 2018. Similarly, the top news story for 2019 so far (2019-12-12) is \textit{AG William Barr's release of his principal conclusions of the Mueller Report}, with an attention score of 22.93 on March 24, 2019.
\end{abstract}

\maketitle

\section{Introduction and background}
It is natural to ask ``what were the top news stories of 2019?'' A partisanship study might ask, ``how often do news stories from different partisan media organizations overlap?'' A retrospective study might ask, ``when did \textit{Hurricane Harvey} begin to receive serious coverage?'', or ``how did the attention given to \textit{Hurricane Harvey} by the media differ from hurricanes that occurred in similar timeframes (but different locations) such as \textit{Irma} or \textit{Maria}?'' Addressing these questions requires the fundamental operation of measuring overlap, or similarity, of news topics across different news sources.

We developed a method of measuring the similarity among news articles in near-real time and quantifying the level of attention the topics in the news stories receive. Specifically, we created a service called \textit{StoryGraph}\footnote{\textcolor{blue}{\url{http://storygraph.cs.odu.edu/}} \& \textcolor{blue}{\url{https://twitter.com/storygraphbot}} } that creates a news similarity graph from 17 left, center, and right news media organizations. StoryGraph quantifies the level of attention the topics in the news stories receive by assigning each an \textit{attention score}. Major breaking news stories are often reported by multiple different news organization within the same time period. Similarly, a major news story is characterized by a high degree of similarity between different pairs of news stories from different news organizations. For example, below is a list of headlines showing a high degree of similarity among news reports collected on October 24, 2018, at 5:34 PM EST from four news organizations, following the incident in which mail bombs were sent to multiple Democratic public figures.
\begin{itemize}[leftmargin=0.5cm]
\item \textbf{Vox}: Explosive devices sent to Clintons, Obamas, CNN: what we know \cite{voxMailBomb}
\item \textbf{FoxNews}: FBI IDs 7 `suspicious packages' sent to Dem figures containing `potentially destructive devices' \cite{foxnewsMailBomb}
\item \textbf{CNN}: Bombs and packages will be sent to FBI lab for analysis \cite{cnnMailBomb}
\item \textbf{Breitbart}: Live Updates: Democratic Leaders Receive Mail Bombs \cite{breitbartMailBomb}
\end{itemize}
The prerequisite for deriving the attention score is calculating the similarity between documents (e.g., news articles). This problem has been studied extensively. Methods that represent documents as vectors \cite{atkins2018measuring, beel2016paper, gabrilovich2007computing} often use the Cosine Similarity vector-based metric to quantify similarity between pairs of documents. Methods that represent documents as sets \cite{tran2014wikipevent, strehl2000impact} often use set-based metrics such as the Jaccard similarity or the Overlap coefficient metric to quantify the similarity between a pair of documents. In this work, we represent each news article as a set of named entities, and utilize a set similarity measure (Section \ref{sec:method}, Step 4) to quantify the degree of similarity between a pair of news documents.

Our investigation into measuring near-real time news similarity and quantifying the attention of news sources has resulted in the following contributions. First, we proposed the attention score, a transparent method for quantifying attention given to a news story by different news sources. The attention score facilitates finding the top news stories for a given day, month, or year. This enabled us to show the top stories of 2018 and 2019 (Table \ref{tab:topNews}). Second, we introduced the StoryGraph service, which has been running for over two years (since August 8, 2017), generating news similarity graphs every 10 minutes from 17 news organizations across the left, center, and right partisanship spectrum. Third, we showed that the StoryGraph service and dataset provides a platform for multiple longitudinal studies (Section \ref{sec:results}). The code for StoryGraph is publicly available \cite{storyGrapher, storyGraphWeb}, and the entire StoryGraph dataset are available upon request.
\setlength{\textfloatsep}{0.1cm}
\begin{table}
   \setlength{\tabcolsep}{1pt}
   \centering
   \caption{List of 17 left (blue), center (purple), and right (red) news media RSS feeds from which StoryGraph extracts news stories. The list of media sources was derived from Faris et al.'s \cite{faris2017partisanship} list of popular media sources.}
   \begin{tabular}{|l|}
          \hline
          \makecell[c]{ \textbf{LEFT} } \\ \hline
          http://www.\textcolor{blue}{politicususa}.com/feed \\ \hline
          https://www.\textcolor{blue}{vox}.com/rss/index.xml \\ \hline
          http://www.\textcolor{blue}{huffingtonpost}.com/section/front-page/feed \\ \hline
          http://www.\textcolor{blue}{msnbc}.com/feeds/latest \\ \hline
          http://rss.\textcolor{blue}{nytimes}.com/services/xml/rss/nyt/HomePage.xml \\ \hline
          http://feeds.\textcolor{blue}{washingtonpost}.com/rss/politics \\ \hline
          \makecell[c]{ \textbf{CENTER} } \\ \hline
          http://rss.\textcolor{purple}{cnn}.com/rss/cnn\_topstories.rss \\ \hline
          http://www.\textcolor{purple}{politico}.com/rss/politics.xml \\ \hline
          http://\textcolor{purple}{abcnews}.go.com/abcnews/topstories \\ \hline
          http://\textcolor{purple}{thehill}.com/rss/syndicator/19109 \\ \hline
          http://feeds.feedburner.com/\textcolor{purple}{realclearpolitics}/qlMj \\ \hline
          \makecell[c]{ \textbf{RIGHT} } \\ \hline
          http://www.\textcolor{red}{washingtonexaminer}.com/rss/news \\ \hline
          http://feeds.\textcolor{red}{foxnews}.com/foxnews/latest \\ \hline
          http://feeds.feedburner.com/\textcolor{red}{dailycaller} \\ \hline
          http://\textcolor{red}{conservativetribune}.com/feed/ \\ \hline
          http://feeds.feedburner.com/\textcolor{red}{breitbart} \\ \hline
          http://www.\textcolor{red}{thegatewaypundit}.com/feed/ \\ \hline
   \end{tabular}
   \label{tab:rss}
\end{table}
\section{Methodology}
\label{sec:method}
\begin{figure*}
    \centering
    \fbox{\includegraphics[width=\textwidth]{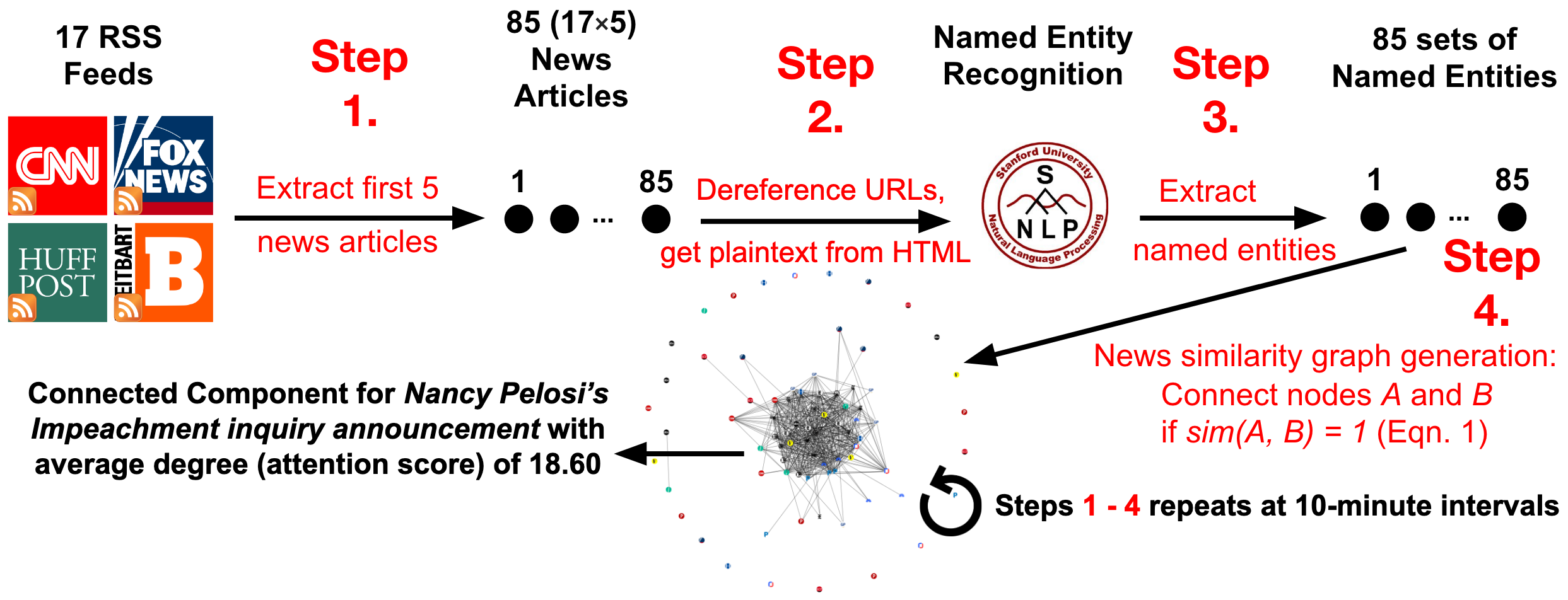}}
    \caption{Overview of StoryGraph illustrating the process of generating a \textit{news similarity graph} is four primary steps.}%
    \label{fig:sgPipeline}%
\end{figure*}
\begin{table*}[h]
  \centering
  \caption{StoryGraph: Worked news similarity (similar news) example. Only \texttt{PERSON} entities are shown here for brevity.}
  \begin{tabular}{|c|c|} \hline 
  \makecell{\textbf{News Article 1}} & \makecell{\textbf{News Article 2}} \\ \Xhline{2\arrayrulewidth}
  \makecell{Christine Blasey Ford's Attorneys Say They Paid\\for Polygraph Test - Breitbart} & \makecell{Live updates: Brett Kavanaugh and Christine Blasey\\Ford hearing on sex assault allegations - CNNPolitics} \\ \hline
  \multicolumn{2}{|c|}{ \textbf{Extracted entities (11 entities common, 40 entities)} } \\ \hline
  
  \makecell{\textbf{blasey}, \textbf{brett}, \textbf{christine}, \textbf{debra}, dianne, \textbf{donald},\\feinstein, \textbf{ford}, hanafin, jerry, \textbf{katz}, \textbf{kavanaugh},\\\textbf{mitchell}, \textbf{rachel}, \textbf{trump}} & \makecell{ashley, banks, ben, \textbf{blasey}, botwinick, \textbf{brett},\\bromwich, \textbf{christine}, cornyn, crapo, cruz, \textbf{debra},\\diamond, don, \textbf{donald}, flake, \textbf{ford}, jeff, jeremy,\\john, \textbf{katz}, \textbf{kavanaugh}, kristina, lee, lisa,\\mcgahn, michael, mike, \textbf{mitchell}, \textbf{rachel}, ryan,\\sasse, sgueglia, ted, \textbf{trump}, zeleny} \\ \hline

  \multicolumn{2}{|c|}{\makecell{ $J($News Article 1, News Article 2$) = \frac{11}{40}$, $O($News Article 1, News Article 2$) = \frac{11}{15}$, $\alpha = 0.30$, \\$Sim($News Article 1, News Article 2$) = 0.30(\frac{11}{40}) + 0.70(\frac{11}{15}) = 0.5958 \geq 0.27 $ (Similar news) }} \\ \hline
  \end{tabular}
  \label{tab:workedExample}
\end{table*}

The StoryGraph process has four steps (Fig. \ref{fig:sgPipeline}) outlined below. First, StoryGraph collects the first five news articles from the RSS feeds of 17 news media organizations (Table \ref{tab:rss}). Second, StoryGraph dereferences the URLs of the news articles and extracts plaintext after removing the HTML boilerplate \cite{nwalaBoilerplateRm}. Third, StoryGraph utilizes the Stanford CoreNLP Named Entity Recognizer \cite{finkel2005incorporating, installCoreNLP} to extract seven entity classes -- \texttt{PERSONS}, \texttt{LOCATIONS}, \texttt{ORGANIZATIONS}, \texttt{DATES}, \texttt{TIME}, \texttt{MONEY}, and \texttt{PERCENT} from the news documents. In addition to these entity classes, we created and extracted text that belong to two additional classes: \texttt{TITLE} and \texttt{TOP-K-TERM}. The \texttt{TITLE} class represents title terms from the news articles, while the \texttt{TOP-K-TERM} class represents the top k (we set $k = 10$) most frequent terms. All text that does not belong to one of the entity classes is discarded. Subsequently, each news article is represented as a set of entities extracted from the article. Fourth, StoryGraph creates a graph where the nodes (set of entities) represent news articles, and an edge between a pair of nodes represents a similarity score beyond some threshold between the nodes (similar news stories). Finally, the attention scores of the connected components of the recently generated graph is calculated. Formally, consider a \textit{news similarity graph G} in which the nodes represent news articles, and an edge between a pair of nodes represents a high degree of similarity (Section \ref{sec:method}, Step 4) between the nodes (similar news stories). Consider the set of $G$'s connected components $C$, such that $\forall c_i \in C$, the nodes (news articles) in $c_i$ originate from multiple news sources. The attention score (Eqn. \ref{eqn:EqnAttnScore}) of a news story represented by a connected component $C_i$ with $|E|$ edges is simply the average degree of the connected component.
\begin{equation}
  \text{Attention score} = \sum_{c \in C_i} deg(c) = \frac{2|E|}{|C_i|}
  \label{eqn:EqnAttnScore}
\end{equation}
\small
\normalsize
\begin{figure*}
    \centering
    \includegraphics[width=\textwidth]{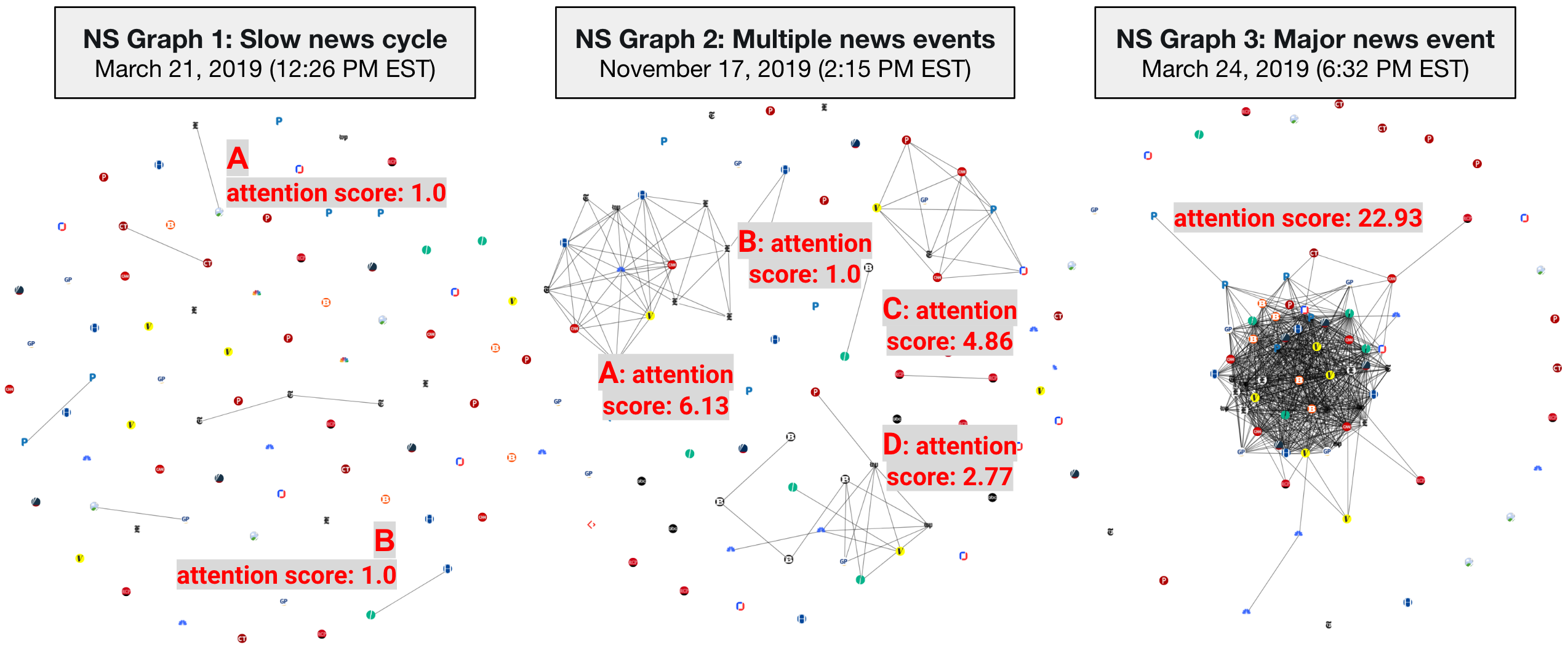}
    \caption{Three News Similarity (NS) graphs illustrating the dynamics of the news cycle. In these graphs, a single node represents a news article, a connected component (multiple connected nodes) represents a single news story reported by the connected nodes. The first (NS Graph 1 \cite{threeGraphs0}) shows what is often referred to as a slow news day; low overlap across different news media organizations resulting in a low attention score (1.0) for news stories (connected components \textit{A} and \textit{B}). The second graph (NS Graph 2 \cite{threeGraphs1}) shows a scenario where the attention of the media is split across four different news stories (connected components \textit{A} -- \textit{D}). The third graph (NS Graph 3 \cite{threeGraphs2}) for the \textit{AG William Barr's release of his principal conclusions of the Mueller Report} story shows a major news event; high degree of overlap/connectivity across different news media organizations, resulting in a high attention score of 22.93}%
    \label{fig:threeGraphs}%
\end{figure*}
\begin{figure*}
    \centering
    \includegraphics[width=\textwidth]{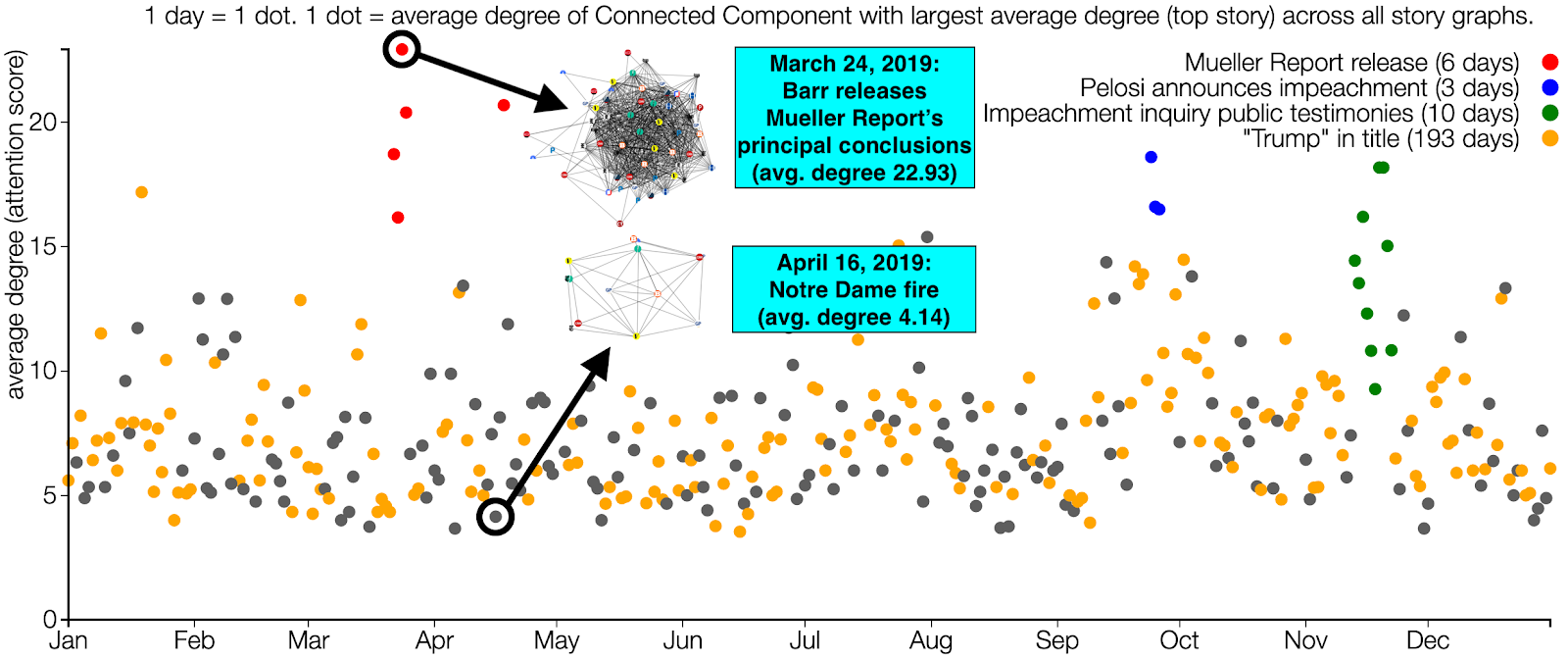}
    \caption{365 dots in 2019 \cite{blog365Dots2019}: Top news stories for 365 days in 2019. Each dot represents the highest attention score across 144 story graphs for a given day.}%
    \label{fig:365DotIn2019}%
\end{figure*}

\begin{figure*}
    \centering
    \subfigure[Kavanaugh and Christine Blasey Ford testify before congress \cite{top2018A} (attention score = 25.85)]{
      \includegraphics[width=.3\linewidth]{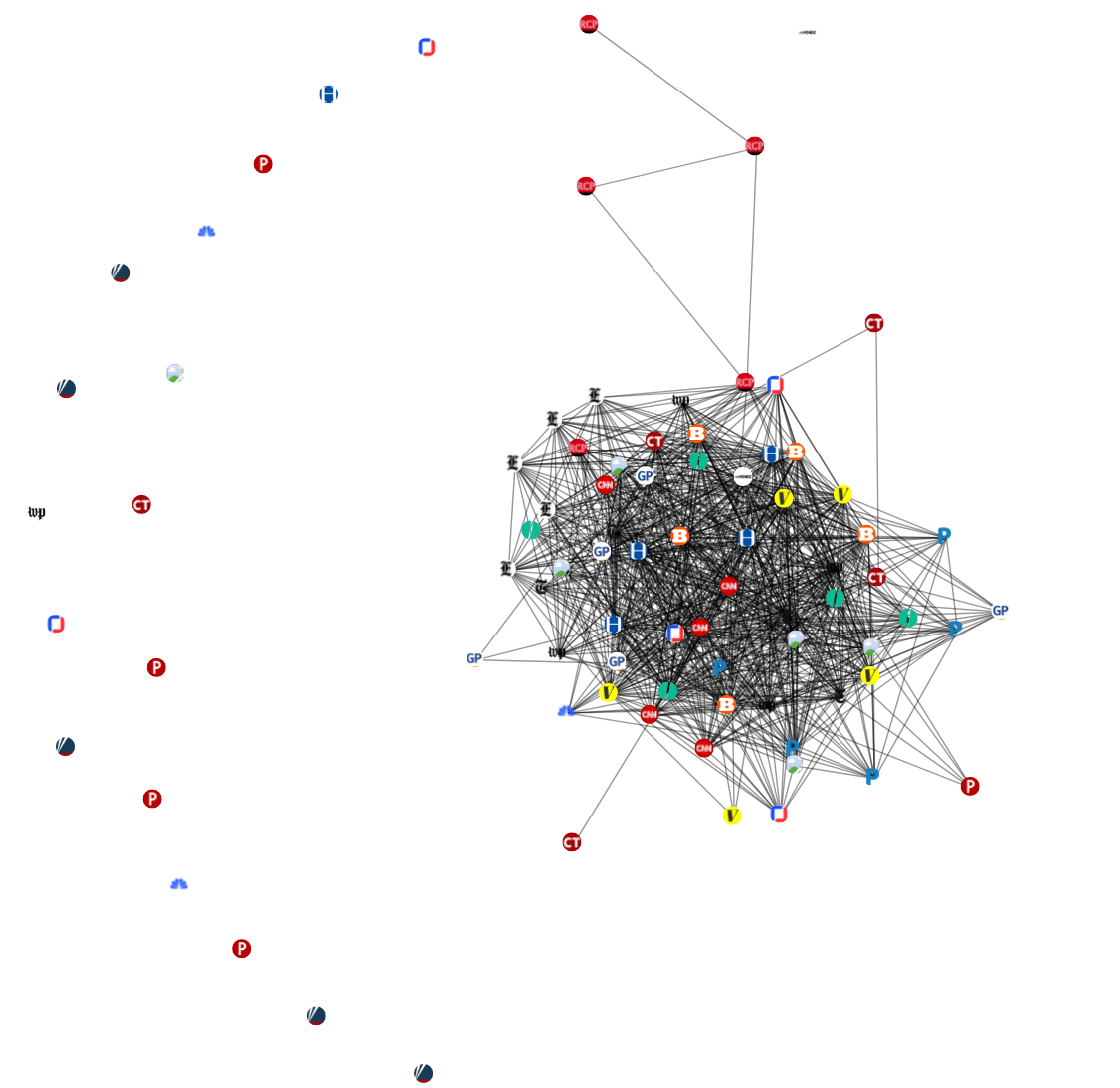}
    }
    \quad
    \subfigure[Nunes memo released \cite{top2018B} (attention score = 18.81)]{
      
      \includegraphics[width=.3\linewidth]{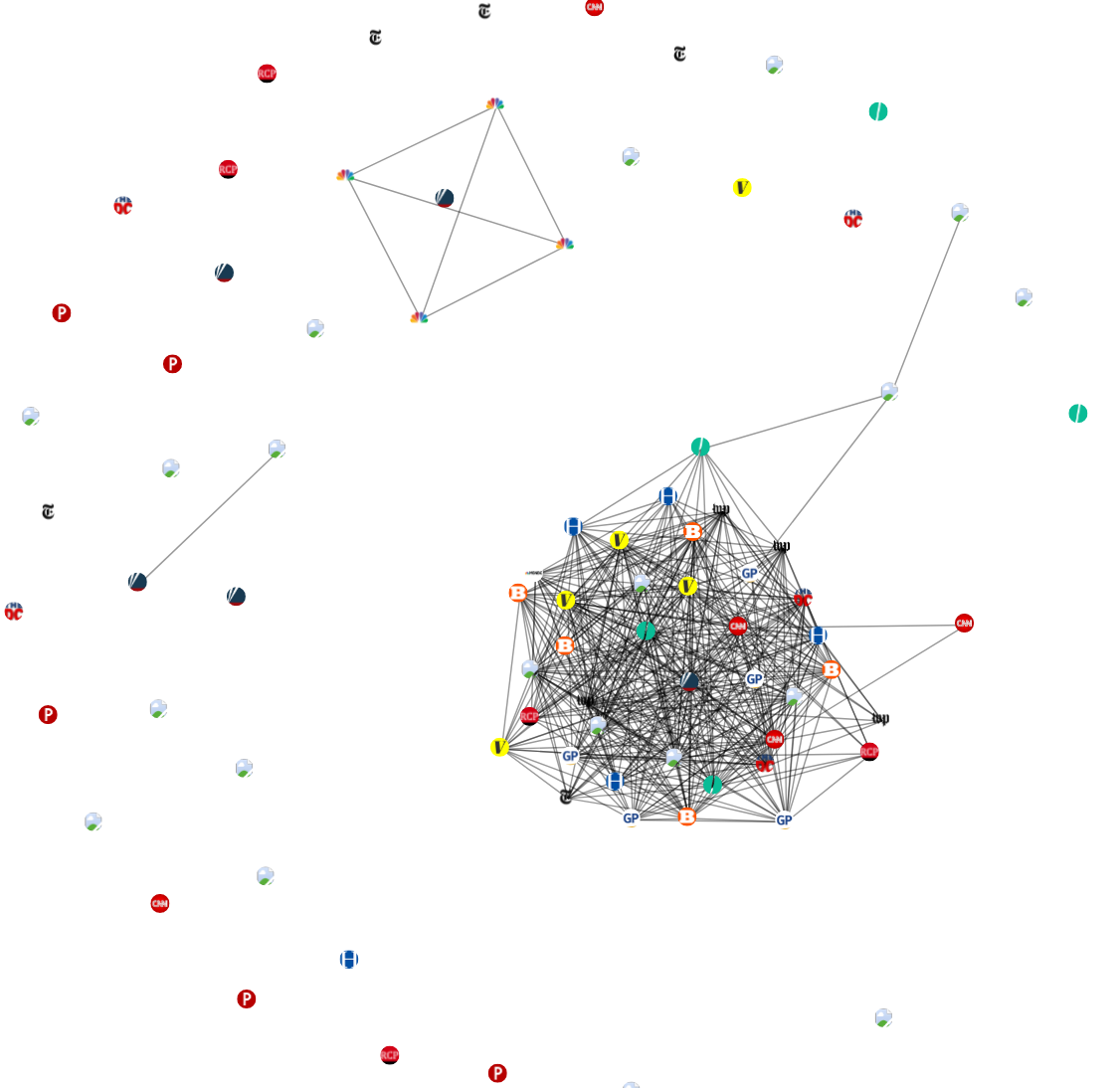}
    }
    \quad
    \subfigure[Trump and Kim Jong Un meet in Singapore \cite{top2018C} (attention score = 18.15)]{
      \includegraphics[width=.3\linewidth]{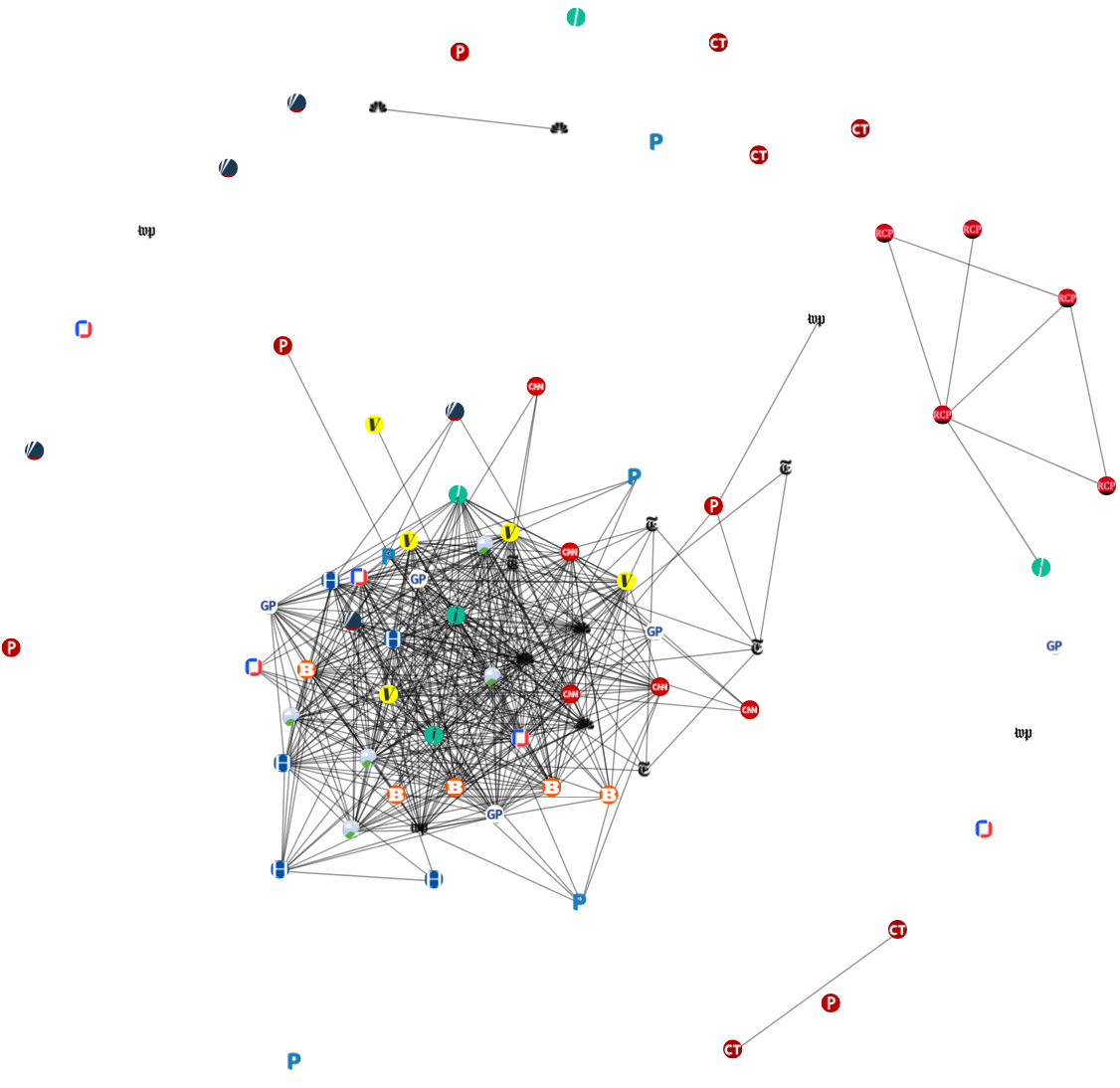}
    }
    \caption{
    StoryGraph: Top three news stories of 2018}
    \label{fig:topNews2018}
\end{figure*}

\begin{figure*}
    \centering
    \subfigure[AG William Barr releases Mueller Report's principal conclusions \cite{top2019A} (22.93)]{
      \includegraphics[width=.3\linewidth]{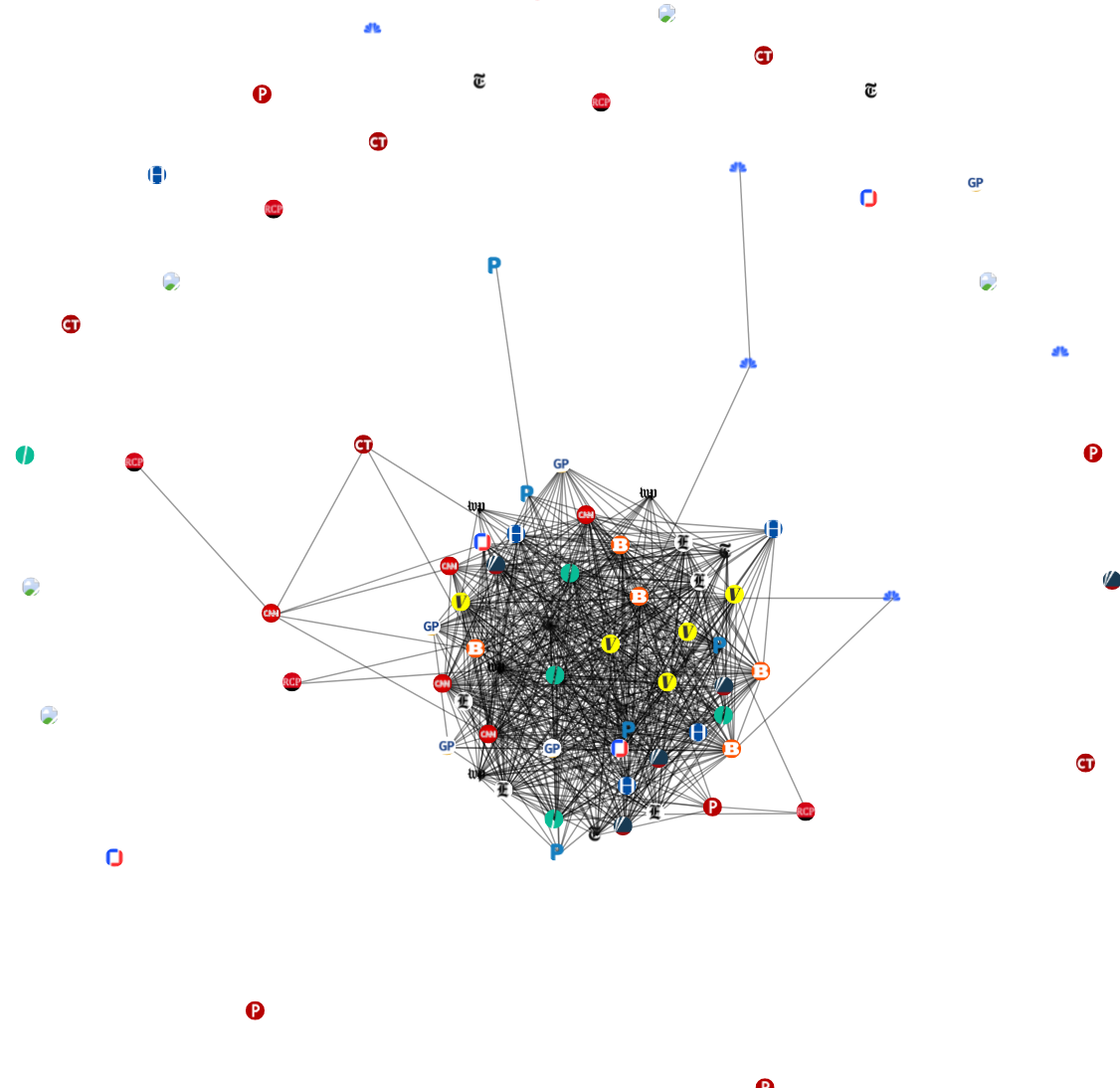}
    }
    \quad
    \subfigure[House Speaker Pelosi announces formal impeachment inquiry \cite{top2019A} (18.60)]{
      
      \includegraphics[width=.3\linewidth]{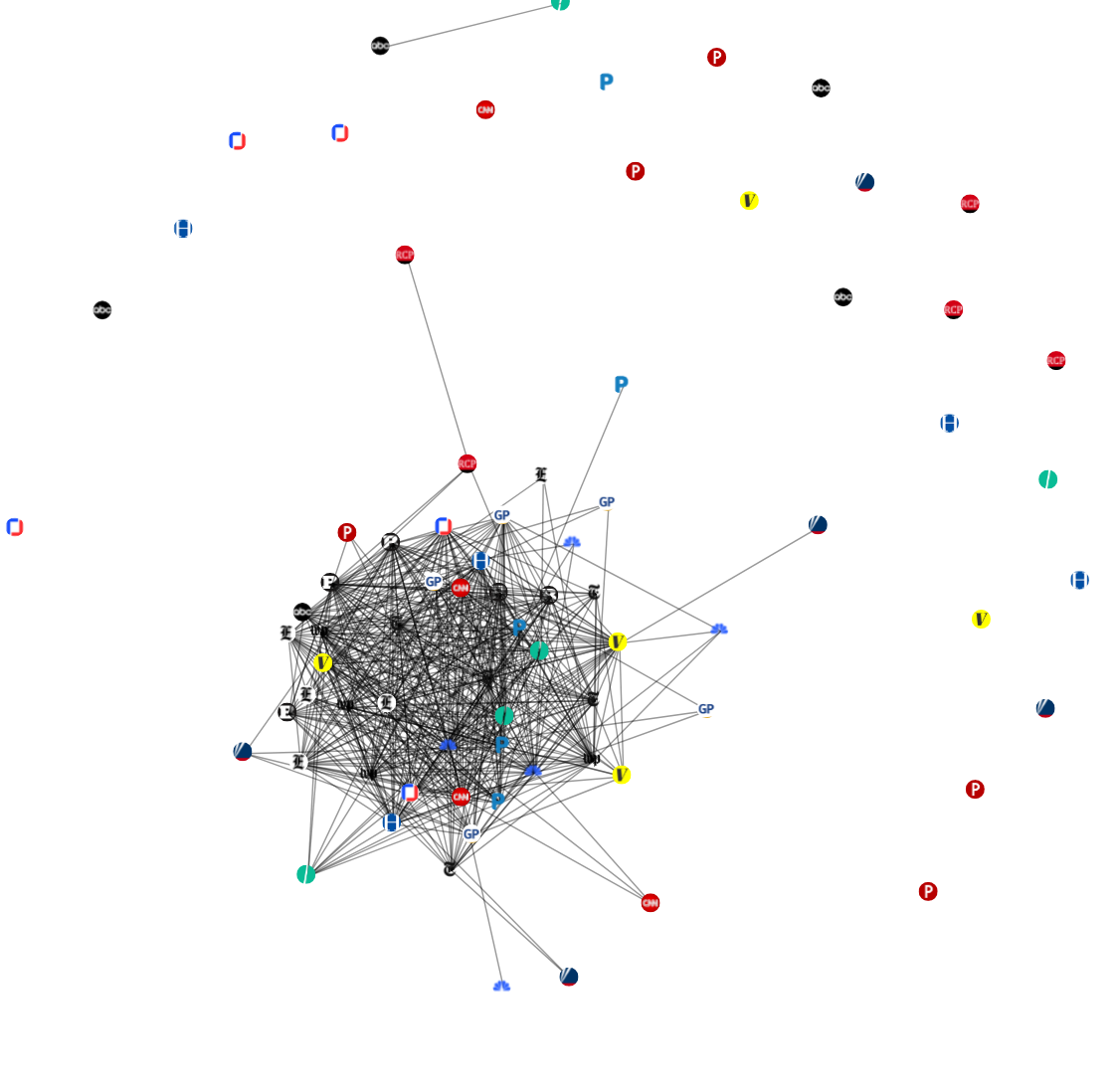}
    }
    \quad
    \subfigure[Impeachment inquiry public testimony \cite{top2019A} (18.18)]{
      \includegraphics[width=.3\linewidth]{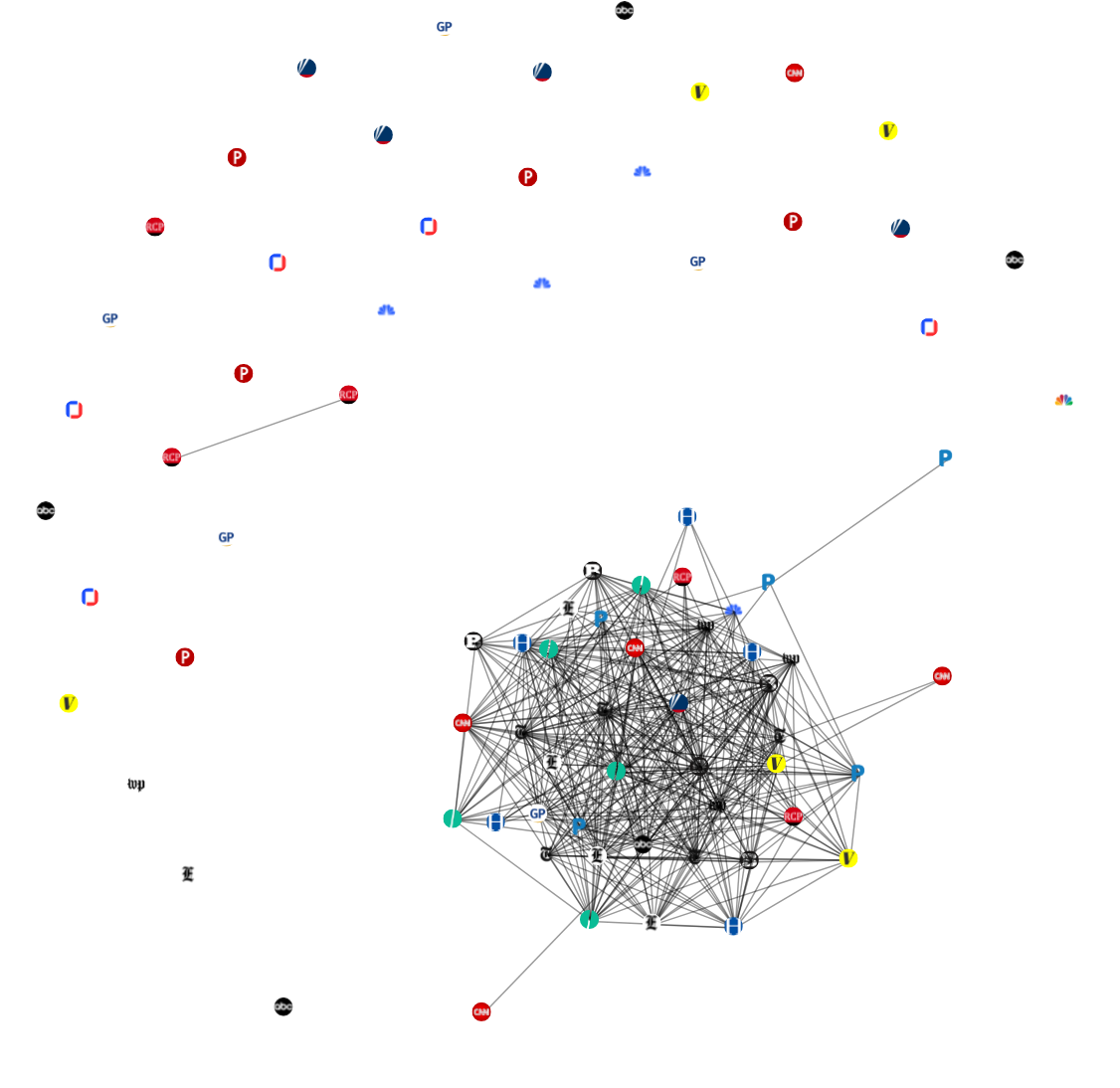}
    }
    \caption{
    StoryGraph: Top three news stories of 2019}
    \label{fig:topNews2019}
\end{figure*}
\setlength{\textfloatsep}{0.1cm}
\begin{table}[h]
  \setlength{\tabcolsep}{1pt}
  \centering
  \caption{StoryGraph: Top news stories of 2018 \cite{blog365Dots2018} \& 2019 \cite{blog365Dots2019} }
  \begin{tabular}{|c|c|c|c|} \hline 
  \textbf{\makecell{Ra-\\nk}} & \makecell{\textbf{Attn.}\\\textbf{score}} & \makecell{\textbf{Date}\\\textbf{(MM-DD)}} & \makecell{\textbf{News story}} \\ \Xhline{2\arrayrulewidth}
  \multicolumn{4}{|c|}{ \textbf{Section 1: Top News Stories of 2018} } \\ \hline
  1 & 25.85 & 09-27 & \makecell{Kavanaugh and Christine Blasey\\Ford testify before congress\\(Fig. \ref{fig:topNews2018}a)} \\ \hline
  2 & 18.81 & 02-02 & \makecell{Nunes memo released\\(Fig. \ref{fig:topNews2018}b)} \\ \hline
  3 & 18.15 & 06-12 & \makecell{Trump and Kim Jong Un\\meet in Singapore\\(Fig. \ref{fig:topNews2018}c)} \\ \hline
  4 & 17.03 & 10-24 & \makecell{Bombs mailed to Clinton, Obama, etc.} \\ \hline
  5 & 16.32 & 03-17 & \makecell{Ex-FBI Deputy Director\\Andrew McCabe fired} \\ \hline 
  \multicolumn{4}{|c|}{ \textbf{Section 2: Top News Stories of 2019} } \\ \hline
  1 & 22.93 & 03-24 & \makecell{AG William Barr releases\\Mueller Report's principal conc.\\(Fig. \ref{fig:topNews2019}a)} \\ \hline
  2 & 18.60 & 09-24 & \makecell{House Speaker Pelosi announces\\formal impeachment inquiry\\(Fig. \ref{fig:topNews2019}b)} \\ \hline
  3 & 18.18 & \makecell{11-19\\11-20} & \makecell{Impeachment inquiry\\public testimony\\(Fig. \ref{fig:topNews2019}c)} \\ \hline
  4 & 17.19 & 01-19 & \makecell{Mueller: BuzzFeed\\Report `Not Accurate'} \\ \hline
  5 & 15.39 & 07-31 & \makecell{2019 Democratic debates} \\ \hline
  \end{tabular}
  \label{tab:topNews}
\end{table}
\setlength{\textfloatsep}{0.1cm}
\subsection{\textbf{Step 1: News article extraction}} StoryGraph extracts the URLs of the first five news articles from each of the 17 RSS feeds (Table \ref{tab:rss}). Next it dereferences each URL yielding 85 HTML documents.
\subsection{\textbf{Step 2: Plaintext extraction}} The HTML boilerplates from the 85 documents from Step 1 are removed \cite{nwalaBoilerplateRm}, yielding 85 plaintext documents.
\subsection{\textbf{Step 3: Named entities extraction}} The 85 plaintext documents from Step 2 are passed into the Stanford CoreNLP Named Entity Recognizer \cite{finkel2005incorporating, installCoreNLP}, yielding 85 different sets of named entities. 
\subsection{\textbf{Step 4: News similarity graph generation}} Given a pair of news articles represented by their respective set of named entities $A$ and $B$, the weighted Jaccard-Overlap similarity $sim(A, B)$ is given by Eqn. \ref{EqnEnitySimMeasure}, where $\beta$ is the coefficient of similarity, defining the threshold two documents must reach to be considered similar ($sim(A, B) = 1$). This threshold was empirically derived from a gold-standard dataset and set to $\beta = 0.27$ and $\alpha = 0.30$. An edge is formed between nodes for which $sim(A, B) = 1$. Table \ref{tab:workedExample} illustrates a simple worked out example.
\begin{equation}
  sim(A, B) = 
  \begin{cases}
       1 & \text{, if $\alpha J(A, B) + (1 - \alpha)O(A, B) \geq \beta$} \\
       0 & \text{, otherwise} \\
  \end{cases}
  \label{EqnEnitySimMeasure}
\end{equation}
$J(A, B)$ is the Jaccard index of both documents, $J(A, B) = \frac{|A \cap B|}{|A \cup B|}$, and $O(A, B)$ is the Overlap coefficient of both documents,\\$O(A, B) = \frac{|A \cap B|}{min(|A|, |B|)}$.\\\\
\noindent
StoryGraph has been running since August 8, 2017, generating a news similarity graph once every 10 minutes. Since then, the application has generated 120,663+ graphs. For a given day, the connected component with the highest average degree (attention score) from 144 candidate graphs maps to the top news story of the day. Similarly, for a given month, the connected component with the highest attention score maps to the top news story for the month. And for a given year, the top \textit{k} news stories is derived by finding \textit{k} connected components with the highest attention scores. Specifically, the top $k$ (e.g., $k = 10$) news stories is the first $k$ connected components from the sorted (in descending order by attention score) list of all news similarity graphs.
\section{Results and Discussion}
\label{sec:results}
We can now quantify ``slow news days'' vs. major news, as well as show that the \textit{Mueller report} was 2019's top story.
\subsection{Slow news day vs. Major news}
Fig. \ref{fig:threeGraphs} illustrates how the attention score (average degree) of the connected components in a news similarity graph helps characterize different news cycle scenarios. All news graphs in this section refer to Fig. \ref{fig:threeGraphs}. Most of the nodes (news articles) in \textit{NS Graph 1} are isolated with few connected components; the news sources mostly report on divergent topics. Consequently, no single news story (e.g., connected component \textit{A} or \textit{B}) receives attention from more than two different news sources.

The second news similarity graph (\textit{NS Graph 2}), unlike the first, shows attention split among four primary news stories. The attention score of each news story represented by the connected component indicates the magnitude of attention given to the news story. For example, connected component \textit{A} (attention score = 6.13) represents the story \textit{Poll: Pete Buttigieg becomes the presidential frontrunner in Iowa - Vox} and \textit{B} (1.0) - \textit{Colin Kaepernick Skips NFL Organized Workout, Wears Shirt Likening Himself to a Slave - Breitbart}.

The third news similarity graph (\textit{NS Graph 3}) indicates a major news event, characterized by a giant connected component (news story) with a high attention score (22.93) indicating a high degree of overlap among news sources. This indicates a scenario when most news sources report on the same story (\textit{AG William Barr's release of his principal conclusions of the Mueller Report}).
\subsection{The top news stories of 2019}
Stories surrounding the release of the Mueller Report (red dots in Fig. \ref{fig:365DotIn2019}, Table \ref{tab:topNews} Section 2, No. 1) received the most attention in 2019. On March 22, 2019, Robert Mueller submitted his report to AG William Barr (attention score = 18.72). Two days later, AG William Barr released his summary (principal conclusions) of the report. This story received the most attention (attention score = 22.93) in 2019. AG William Barr's principal conclusions of the Mueller report was received with skepticism by the Democrats who claimed the conclusions were highly favorable to President Trump. In contrast, the Republicans claimed the summary exonerated the President from any wrongdoing. The next top story in 2019 (blue dots in Fig. \ref{fig:365DotIn2019}, Table \ref{tab:topNews}, Section 2, No. 2) with attention score of 18.60 was Speaker Nancy Pelosi's announcement of an official impeachment inquiry (September 24, 2019) four days after the whistleblower's report. Similarly, at rank three (green dots in Fig. \ref{fig:365DotIn2019}) were stories chronicling the public testimonies of the impeachment inquiry.
\section{Future Work and Conclusion}
\label{sec:conclusion}
StoryGraph has been generating one news similarity graph every 10 minutes since August 2017. A single graph file includes the URL of the news articles, plaintext, entities, publication dates, etc. In this paper, we only reported the result of two studies. The first studies the dynamics of the news cycle (\textit{slow news cycle} vs. major news event). The second utilized attention scores to facilitate finding top stories.

StoryGraph provides the opportunity for further study beyond the two presented here. For example, a study focused on the coverage of mass shootings can utilize StoryGraph to approximate how much attention the 2018 Parkland, Florida shooting received compared to the 2019 Dayton, Ohio and El Paso, Texas mass shootings. A different study could narrowly apply news similarity to focus on a single news organization, e.g., FoxNews, in order to identify the news stories where they focus the most attention, or compare the attention span of different events. Therefore, we believe the StoryGraph process of quantifying news similarity and the attention of news sources provides a valuable means for studying news.
\begin{acks}
This work was supported in part by IMLS LG-71-15-0077-15. This is an extended version of the paper accepted at Computation + Journalism Symposium 2020, which has been postponed because of COVID-19. We also appreciate the help of Sawood Alam in the deployment of StoryGraph.
\end{acks}

\bibliographystyle{ACM-Reference-Format}
\bibliography{NwalaSG}

\end{document}